# Probing dynamics of water mass transfer in organic porous photocatalyst water-splitting materials by neutron spectroscopy

Mohamed Zbiri[1]*, Catherine M. Aitchison[2], Reiner S. Sprick[2], Andrew I. Cooper[2], Anne A. Y. Guilbert[3]*

[1]Institut Laue-Langevin, 71 Avenue des Martyrs, Grenoble Cedex 9 38042, France

[2]Department of Chemistry and Materials Innovation Factory, University of Liverpool, Crown Street, Liverpool L69 7ZD, U.K.

[3]Department of Physics and Centre for Plastic Electronics, Imperial College London, Prince Consort Road, London SW7 2AZ, U.K.

**ABSTRACT:** The quest for efficient and economically accessible cleaner methods to develop sustainable carbon-free energy sources induced a keen interest in the production of hydrogen fuel. This can be achieved via the water-splitting process exploiting solar energy but requiring the use of adequate photocatalysts to reach this goal. Covalent triazine-based frameworks (CTFs) are potential target photocatalysts for water-splitting. Both electronic and structural characteristics of CTFs, particularly energy levels, optical bandgaps and porosity directly relevant for water-splitting, can be engineered through chemical design. Porosity can, in principle, be beneficial to water-splitting by providing larger surface area for the catalytic reactions to take place. However, porosity can also affect both charge transport within the photocatalyst and mass transfer of both reactants and products, thus impacting the overall kinetics of the reaction. Here, we focus on the link between chemical design and water (reactants) mass transfer, playing a key role in the water uptake process and the subsequent hydrogen generation in practice. We use neutron spectroscopy to study the mass transfer of water in two porous CTFs, CTF-CN and CTF-2, that differ in the polarity of their struts. Quasi-elastic neutron scattering is used to quantify the amount of bound water and the translational diffusion of water. Inelastic neutron scattering measurements complement the quasi-elastic neutron scattering study and provides insights into the softness of the CTF structures and the changes in librational degrees of freedom of water in the porous CTFs. We show that two different types of interaction between water and CTFs take place in CTF-CN and CTF-2. CTF-CN exhibits smaller surface area and water uptake due to a softer structure than CTF-2. However, the polar cyano group interacts locally with water leading to a large amount of bound water and a strong rearrangement of the water hydration monolayer while the water diffusion in CTF-2 is principally impacted by the micro-porosity. The current study leads to new insights into the structure-dynamics-property relationship of CTF photocatalysts that pave the road for a better understanding of the guest-host interaction at the basis of water splitting applications.

## INTRODUCTION

Hydrogen has been suggested as the energy carrier of the future as it can be stored and does not emit greenhouse gases at the point of use.[1] However, most hydrogen is still produced using steam reforming processes emitting large amounts of carbon dioxide when produced. The generation of hydrogen using clean methods have, therefore become an area of intense research. Photocatalytic water splitting has been of particular interest as it uses water and solar light, which are abundant on Earth surface.[2-5]

In the process, a catalyst is used to generate charge-carriers that facilitate water reduction and oxidation. The catalysts are typically inorganic semiconductors.[2,3] Tremendous progress has been made in recent years in terms of improving the catalyst materials and the overall systems.[1,6] Nevertheless, organic materials have received significant attention over the last decade as they can be synthesized using many potential building blocks. Hence, the properties of the materials can be tuned by chemical design.[4,7]

Particularly, carbon nitrides have been studied as organic photocatalysts. Carbon nitrides are usually made through high-temperature condensation reactions and, exhibit, as a result, many defects including end-groups.[8,9]

Linear conjugated polymers,[10–16] conjugated microporous polymers (CMPs),[16–24] covalent triazine-based frameworks (CTFs)[25–28] and covalent organic frameworks[29–33] have been extensively studied in recent years allowing for fine-tuning of material properties. Several factors have been identified to be necessary for the activity of polymer photocatalysts, such as the ability of a photocatalyst to absorb light,[13,21,23] the position of the redox potentials,[13] exciton separation,[14] crystallinity,[29] and wetting of the surface.[15,17,20] Sufficient driving forces for both half-reactions, numbers of absorbed photons and dispersibility in the reaction mixtures are necessary properties but not sufficient conditions for a linear conjugated polymer or a CTF to be an active photocatalyst. The two-best performing CTFs among 44 synthesized CTFs were CTFs with polar groups, cyano group and sulfone group, in their linkers.[26] Previously, modelling suggested that adding sulfone groups increased the polarity of the local environment of the photocatalyst and thus, improved the thermodynamic driving force for the oxidation.[34,35] However, for both families of materials, the hydrogen evolution rates, at least when used with triethylamine as hole scavenger, were more limited by thermodynamics driving forces for proton reduction than by light absorption or the oxidation of the hole scavenger. Indeed, electron affinity and dispersibility were found for CTFs to be the dominant variables.[26] Better dispersibility leads to higher interfacial surface area between the photocatalysts and the reaction mixture. In the case of porous photocatalysts, such as covalent organic frameworks (COFs) and CMPs, high Brunauer-Emmett-Teller surface areas ($SA_{BET}$) have been suggested to be partly responsible for their enhanced photocatalytic activity as active sites within the material can potentially be accessed by water.[18,29,36,37] However, in a previous paper, we studied CMPs and their linear polymer analogues and we found that the porous materials do not always outperform their non-porous analogues.[36] Beyond the thermodynamic properties such as driving forces, one should not forget about the kinetics of the photocatalytic water-splitting process. Microporosity is likely to reduce the charge transport rates in the solid state and constrain the diffusion of both reactants and products in the pore. Therefore, it is important to gain a deeper understanding of the quantitative relationship between chemical structure and mass transfer of reactants in organic porous photocatalysts.

Measurements, such as water uptake do not provide any information about water dynamics and crucially, no information about interaction of water with the surface of the materials at the molecular level can be obtained. Organic materials are made mainly of light elements. Neutrons, unlike X-ray, do not discriminate atomic species as a function of their size. Neutrons are, in particular, very sensitive to hydrogens and thus, deuteration can be used for a contrast variation purpose between the photocatalyst and water (Table 1). Neutrons are therefore a useful probe to study both organic photocatalysts and water. Neutron spectroscopy allows probing length scales and time scales relevant to the atomic and molecular interactions, covering the microscopic guest-host dynamics that takes place on the picosecond and nanometer scales. Thus, neutron spectroscopy is a useful technique to understand the dynamics and intersection of water guest molecules in organic photocatalytic porous host materials. In this context, quasi-elastic neutron scattering (QENS) and inelastic neutron scattering (INS) have been applied recently to map in details the microstructural dynamics up to the nanosecond of the conjugated polymer poly(3-hexylthiophene), under both its regioregular and regiorandom forms.[38] QENS has been used to study diffusion pathways and relaxation timescales of lithium ions in inorganic battery active materials,[39] methane diffusion in metal organic frameworks,[40] and the rotational dynamics of hydrogen adsorbed in covalent organic frameworks.[41] QENS has also been used to study the state of water when interacting with oligonucleotide crystals[42] and porous organic cages[43] showing distinct differences between bulk water and confined water interacting with the crystal surface. We previously studied water penetration and dynamics for a polar dibenzo[b,d]thiophene sulfone and a fluorene CMP using QENS in the context of photocatalysis,[36] however, the materials in the study were vastly different in terms of their polarities and water uptake.

Here, we study the influence of the network structure on the dynamics of water transfer within the networks of CTFs by means of neutron spectroscopy (QENS and INS). QENS enables us to study the underlying dynamics of water within the networks and INS allows us to draw conclusions about interactions of water with the network at the molecular level. Water can exhibit different degrees-of-freedom within a host material, and as a such free water can undergo a transition to a constrained (or trapped) state, and/or can be of a bonding nature. We select as a model system the previously reported CTF-2,[25] which contains an apolar biphenyl-linker, and CTF-CN,[26] which contains a polar cyano-group in the linker. Importantly, CTF-2 and CTF-CN present similar properties, like electron affinity, ionization potentials, optical bandgap and both swell.[25,26] This allows us to study two structurally related porous materials that differ in the length and polarity of their struts.

## RESULTS

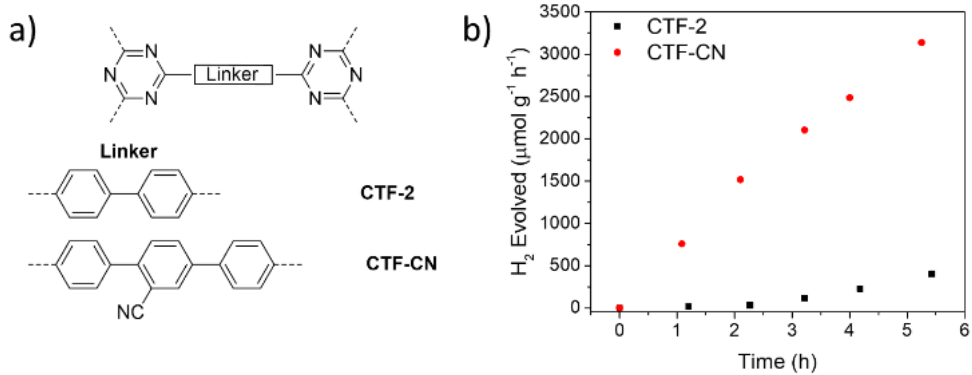

FIGURE 1. (a) Chemical structures of CTF2 and CTF-CN. (b) Hydrogen evolution of CTF-2 and CTF-CN. Linear hydrogen evolution rates were determined to be 118 µmol h$^{-1}$g$^{-1}$ for CTF-2 and 595 µmol h$^{-1}$g$^{-1}$ for CTF-CN. Conditions: Photocatalyst (25 mg loaded with 3 wt. % Pt by photodeposition of H$_2$PtCl$_6$) suspended in water/TEA (95:5 vol. %, 25 mL) illuminated by a 300 W Xe light source fitted with a λ > 420 nm filter.

Both CTF-2[25] and CTF-CN[26] (Figure 1) were made using previously reported methods and tested as photocatalysts for hydrogen production from water in the presence of triethylamine[25,26]. Platinum was added as a co-catalyst via photodeposition from H$_2$PtCl$_6$ and linear rates of 118 µmol h$^{-1}$ g$^{-1}$ for CTF-2 and 595 µmol h$^{-1}$ g$^{-1}$ for CTF-CN were determined (Figure 1b). Residual palladium from the synthesis step also acts as a co-catalyst. Note that, as reported previously,[25,26] even without adding platinum, a significant hydrogen production from both CTF-2 and CTF-CN can still be observed. Both materials were found to be porous to nitrogen with SA$_{BET}$ determined to be SA$_{BET}$ = 873 m$^2$ g$^{-1}$ for CTF-2 and 548 m$^2$ g$^{-1}$ for CTF-CN (Figure 2a). The higher SA$_{BET}$ observed for CTF-2, although having a shorter linker, is in line with the previous study where CTF-2 was found to have a higher SA$_{BET}$ than CTF-1 (one phenylene) and CTF-3 (3 phenylenes). [25] The relatively high SA$_{BET}$ of both materials together with the hydrophilic triazine building block in the materials may allow for water penetration into the network. The pore size distribution, extracted from the nitrogen sorption isotherm presented in Figure 2 using the Barrett-Joyner-Halenda (BJH) method,[44] shows the presence of micropores for both CTF-2 and CTF-CN, and the additional presence of mesopores in the case of CTF-CN only (Supporting Information, Figure S3). Micropores in CTF-2 are found to be slightly wider than in CTF-CN. Interestingly, from water sorption measurements (Figure 2b), CTF-2 shows a higher water uptake than CTF-CN. Water sorption seems to be linked with the higher surface area of CTF-2 rather than to the increase polarity of CTF-CN, although condensation due to the interaction of water with the surface cannot be ruled out. Further structural and optical characterization of both materials can be found in references 25 and 26.

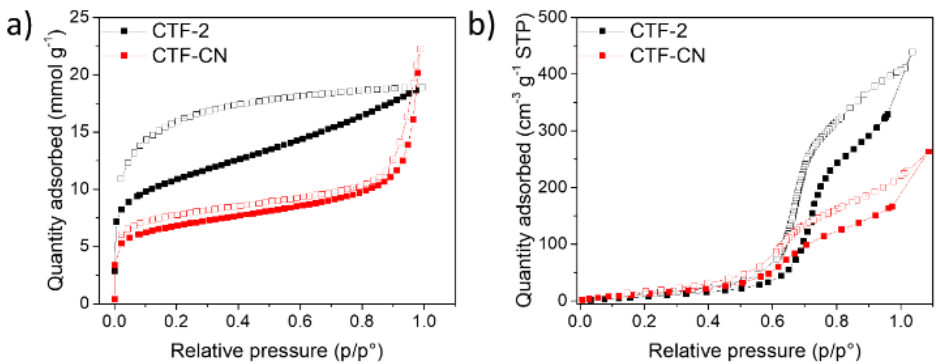

FIGURE 2. (a) Nitrogen sorption isotherm for CTF-2 and CTF-CN measured at 77.3 K and up to 1 bar (desorption curves shown as open symbols). (b) Water uptake isotherms for CTF-2 and CTF-CN measured at 293.15 K and up to 23.393 mbar (desorption curves shown as open symbols).

We probe the molecular diffusivity of water at the surface/within the pores of the CTFs, to quantify the internal mass transfer of water in this systems, using QENS. As stated above, significant hydrogen production from CTFs was observed without adding platinum.[25,26] Platinum has a non-negligible neutron absorption cross-section, weakening the scattered signal and could additionally add a further complication to the analysis and interpretation of the neutron spectroscopy data. The aim of the present study is to focus on the dynamics of CTFs and water, as the main and unique components of the guest-host system. The CTFs were therefore not loaded with platinum for the present neutron scattering study. The QENS study, offering insights into the local guest-host dynamics, is underpinned by INS to gain further insights, vibrationally, into the water interaction with the CTFs leading to the transition from free water to constrained and bound water. We used two neutron incident wavelengths (5 Å and 8 Å) to enhance the QENS resolution (8 Å) at low Q and to cover an extended Q-range (5 Å). Indeed, the use of the 5 Å setting gives a resolution at the elastic line of ∼ 0.1 meV and a Q-range of 0.2-2.3 Å$^{-1}$, while 8 Å leads to an improved resolution of 0.03 meV, but a limited Q-range of 0.1-1.3 Å$^{-1}$. This should enable capturing suitably both the translational and rotational motions of water.

Table 1 gathers the neutron incoherent cross section of the different systems presently studied, highlighting how neutron spectroscopy is very sensitive to hydrogens, and the usefulness of using deuterated water (D$_2$O) to enable further tuning the contrast between the CTFs (host) and water (guest), and potentially revealing the impact of water on the CTFs molecular motions.

TABLE 1. Neutron incoherent cross section (cm$^{-1}$) of the samples studied in this work. The density of the CTFs is taken to be 0.8 g cm$^{-3}$.

| Water concentration (wt. %) | CTF-CN | CTF-2 | H$_2$O | D$_2$O |
|---|---|---|---|---|
| 0.0 | 1.707 | 1.933 | | |

| | | | | | | |
|---|---|---|---|---|---|---|
| **100.0** | | 5.621 | 0.138 | **33.3** | 0.644 | 3.749 |
| **21.0** | 0.358 | 4.441 | | **35.7** | 0.690 | 0.089 |
| **20.6** | 0.352 | | 0.110 | | | |

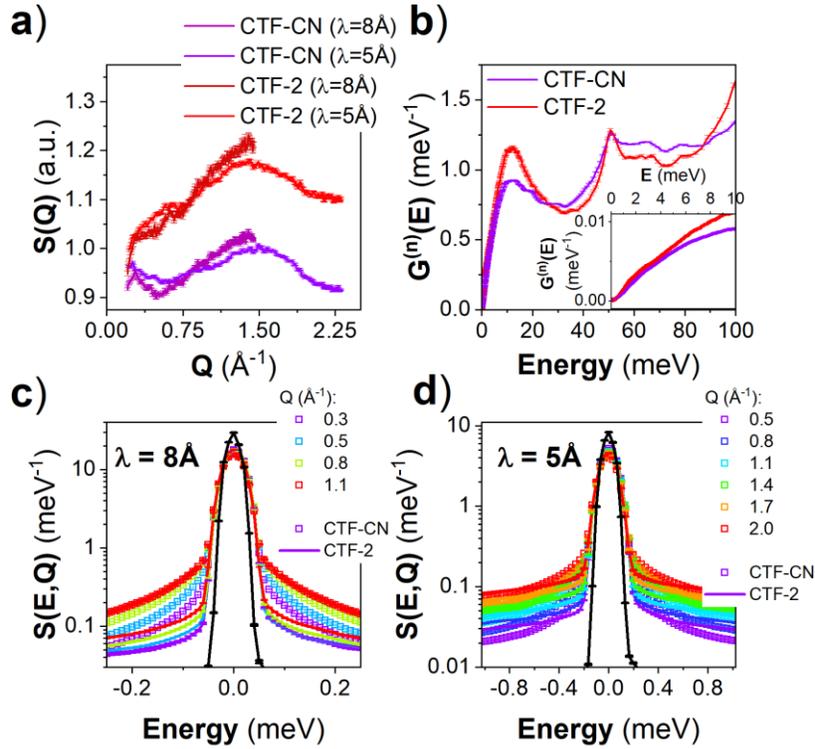

FIGURE 3. (a) Neutron diffractograms of CTF-CN and CTF-2 extracted from the QENS measurements using two neutron incident wavelengths 5 Å and 8 Å (two different instrumental energy/time resolutions). (b) Area-normalized generalized phonon density of states (GDOS)[45] of CTF-CN and CTF-2 using a neutron incident wavelength of 5 Å. The inset shows the evolution of the Debye growth (0-10 meV region). Area-normalized Q-dependent QENS spectra of dried CTF-CN and CTF-2 using the two wavelengths (c) λ = 8 Å and (d) λ = 5 Å. The instrumental resolution function from a vanadium sample is shown in (c) and (d) as the narrow black solid elastic line.

Figure 3 presents the neutron diffractograms, the generalized phonon density of states (GDOS)[45] and the QENS spectra of the dried CTF-CN and CTF-2.

In line with their powder X-ray diffraction patterns,[25,26] both CTFs appear rather amorphous with a broad diffraction peak at around 1.5 Å$^{-1}$ (Figure 3 (a)). CTF-CN seems to exhibit another Bragg peak at low Qs, ~ 0.3 Å$^{-1}$. The GDOS spectra (Figure 3 (b)) exhibit similar features, both in terms of phonon bands and the slope of the Debye growth (0-10 meV region, inset Figure 3(b)), for both CTFs. These features are more pronounced in the case of CTF-2, pointing towards a stiffer and more frustrated aspect of the structure of CTF-2. Moreover, the Debye growth of CTF2 seems to be slightly red-shifted with respect to that of CTF-CN, which points towards CTF-2 being more disordered than CTF-CN. Although the QENS spectra feature some resemblance with a strong elastic contribution, CTF-CN exhibits noticeable dynamics within the covered energy range, either at 8 Å (Figure 3 c) and 5 Å (Figure 3 d). From these measurements, it is clear that the smaller linker of CTF-2 with two benzene rings leads to a stiffer structure than CTF-CN as observed by INS and that the relatively higher softness of CTF-CN leads to a more pronounced dynamics, as observed in the QENS data.

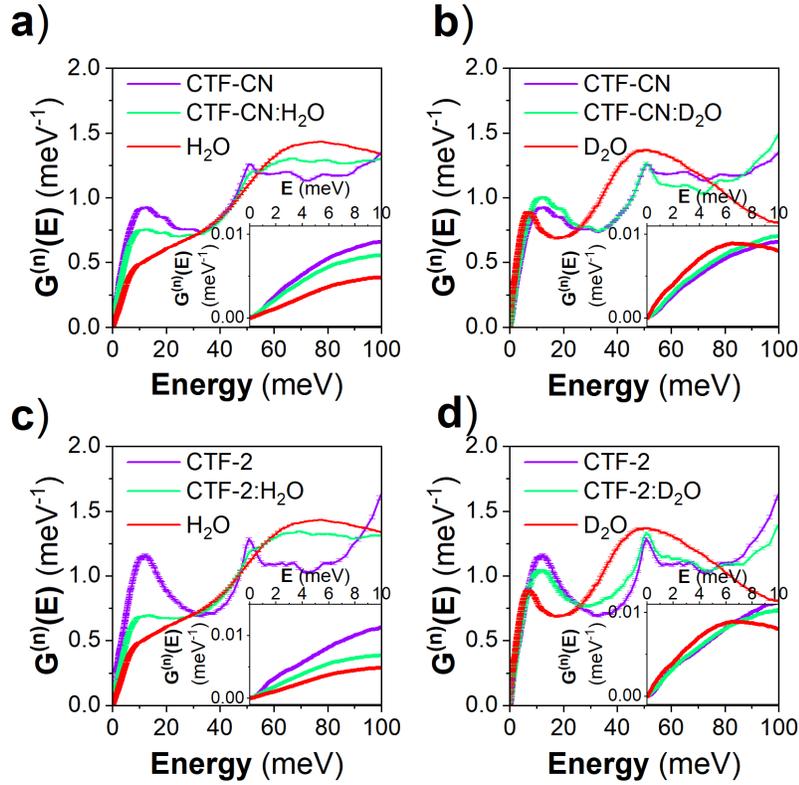

FIGURE 4. Area-normalized generalized phonon density of states (GDOS), from measurements at 5 Å, of: (a) dried CTF-CN, wetted CTF-CN with $H_2O$ and $H_2O$, (b) dried CTF-CN, wetted CTF-CN with $D_2O$ and $D_2O$, (c) dried CTF-2, wetted CTF-2 with $H_2O$ and $H_2O$, and (d) dried CTF-2, wetted CTF-2 with $D_2O$ and $D_2O$. The insets show the evolution of the Debye growth (0-10 meV region).

Figure 4 shows the GDOS spectra of CTF-CN and CTF-2, both dried and mixed with either $H_2O$ or $D_2O$. Interestingly, almost no differences are observed in the GDOS of CTF-2 upon addition of $D_2O$ while the changes in GDOS spectra of CTF-CN cannot be simply explained by a neutron weighted average (concentration, neutron scattering cross sections) of the spectra of the CTF and $D_2O$, especially for the strength of the peak around 10 meV. Moreover, a softening of the structure of CTF-CN upon hydration is reflected by a red shift of the slope of the Debye growth (energy range 0 - 10 meV, insets of Figure 4), but not observed for wetted CTF-2. This points towards the occurrence of specific interactions between CTF-CN and water leading to a further ease of accommodation of water in the CTF-CN structure, which could probably be due to the presence of the -CN group.[26] The GDOS in that case are dominated by the $H_2O$ signal and no clear indication of interactions can be seen without further analysis.

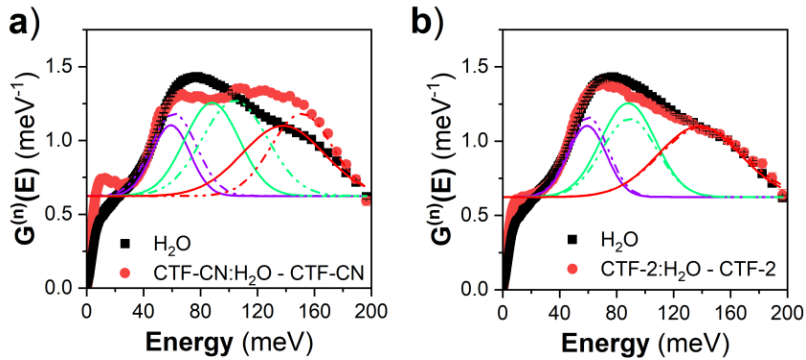

FIGURE 5. Area-normalized generalized phonon density of states (GDOS) of bulk reference $H_2O$ and $H_2O$ in (a) CTF-CN and (b) CTF-2, from measurements at 5 Å. The GDOS of $H_2O$ in the CTF samples is presented as the difference of the wetted CTFs (either CTF-CN:$H_2O$ or CTF-2:$H_2O$) and dried CTFs (either CTF-CN or CTF-2, respectively). The broad peak around 80 meV is assigned to the libration of water and is fitted with a combination of 3 Gaussians representing the rock, wag, and twist modes of water.[46,47] The solid lines are the fits for bulk $H_2O$ and the dotted lines are fits for the $H_2O$ in the CTF.

Insights into the specific behavior of water in the two CTFs can be gained by exploiting the GDOS of water and wetted CTFs. Figure 5 shows the GDOS of $H_2O$ as a "reference", compared to the difference of the GDOS of the wetted CTFs(CTF-2:$H_2O$ or CTF-CN:$H_2O$), and dried CTFs (CTF-2 or CTF-CN, respectively). The differences are obtained as follow:

$$\Delta G^{(n)}(E) = \frac{\left(\frac{\sigma_{CTF}}{m_{CTF}} + \frac{\sigma_{H_2O}}{m_{H_2O}}\right)_{CTF:H_2O} \times G^{(n)}_{CTF:H_2O} - \left(\frac{\sigma_{CTF}}{m_{CTF}}\right)_{CTF} \times G^{(n)}_{CTF}}{\left(\frac{\sigma_{CTF}}{m_{CTF}} + \frac{\sigma_{H_2O}}{m_{H_2O}}\right)_{CTF:H_2O} - \left(\frac{\sigma_{CTF}}{m_{CTF}}\right)_{CTF}}$$

where $\sigma$ is the neutron cross-section, $m$ is the mass of the material in the sample and $G^{(n)}(E)$ the measured area-normalized GDOS. The differences in GDOS can be thus seen as the signal of $H_2O$ in the CTFs. The broad peak around 80 meV for bulk $H_2O$ is assigned to the libration of water and is fitted with 3 Gaussians (solid lines) representing the rock, wag and twist modes of water.[46,47] We fit similarly the signal of $H_2O$ in the CTFs (doted lines). The comparison of the fits of bulk $H_2O$ and $H_2O$ in the CTF Gaussian-wise clearly highlights a pronounced hindrance and change in the vibrational distribution of the librational degrees-of-freedom of $H_2O$ in CTF-CN as compared to $H_2O$ in CTF-2. Furthermore, the intensity of the low-energy feature of water, at approximately 10 meV, increases when $H_2O$ is in both CTFs although more significantly for CTF-CN. The hindrance of the librational degrees-of-freedom of water in CTF-CN is a clear indication of the transition from free water to constrained water and/or bound water, and this will be further quantified below. The changes in the low-energy features could reflect a change in the organization of water, especially in the hydration monolayer of both CTFs, with a more pronounced effect in CTF-CN.

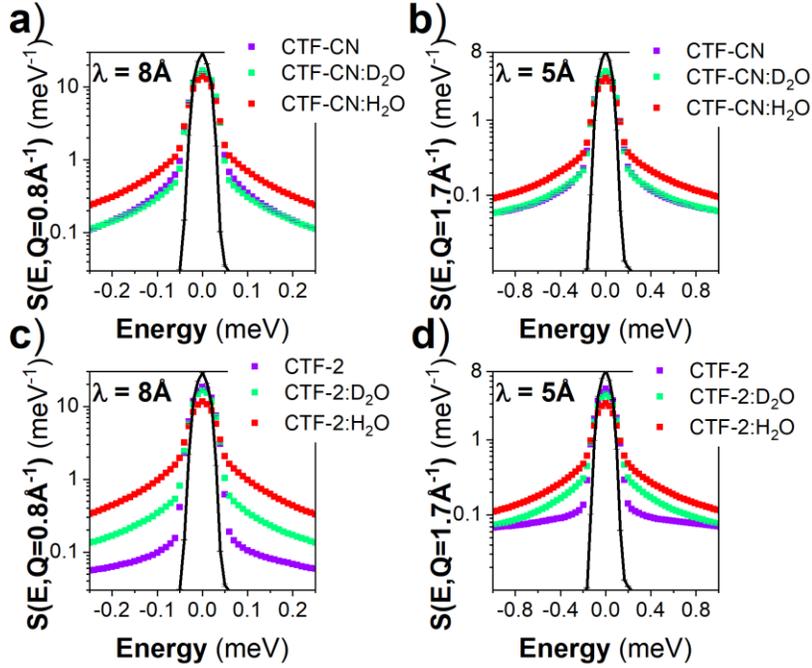

FIGURE 6. Area-normalized QENS spectra of (a,b) dried and wetted (with $H_2O$ or $D_2O$) CTF-CN and (c,d) dried and wetted (with $H_2O$ or $D_2O$) CTF-2, using two neutron incident wavelengths 8 Å (a,c) and 5 Å (b,d), ensuring two different instrumental energy/time resolutions. The instrumental resolution function from a vanadium sample is shown as the narrow black solid elastic line.

Figure 6 shows the QENS spectra of CTF-CN and CTF-2, both dried and mixed with either $H_2O$ or $D_2O$. When $D_2O$ is added, no noticeable differences, as compared to the dried case, are observed in the QENS spectra for CTF-CN while a broadening is observed in the case of CTF-2. In both cases, the expected contribution to the total QENS spectra from the CTF material is supposed to be larger than from $D_2O$ although to a lesser extent for CTF-CN (Table 1), as stemming from the difference in the incoherent neutron cross-section. However, due to the differences in types of contribution i.e. CTFs have mainly an elastic contribution while $D_2O$ has mainly a quasi-elastic contribution, nothing can be concluded without further analysis. For CTF-CN, a slight narrowing of the QENS signal is observed at the highest incident wavelength (8 Å), indicating potentially a further frustration of the structure when water is present. When $H_2O$ is added to the CTFs, in both cases, a strong broadening of the QENS signals is observed. In this case, the quasi-elastic contributions are dominated by $H_2O$.

The QENS signals can be fitted to extract quantitative information about the mass transfer of water within the present CTF materials and also to estimate the amount of bound water in both CTFs. The dynamic structure factor of water $S_{water}(Q,\omega)$ can be expressed as a convolution of the dynamic structure factors of the vibrational $S_V(Q,\omega)$, translational $S_T(Q,\omega)$ and rotational motions of water $S_R(Q,\omega)$.[48]

$$S_{water}(Q,\omega) = S_V(Q,\omega) \otimes S_T(Q,\omega) \otimes S_R(Q,\omega)$$

$S_V(Q,\omega)$ can be written as: $S_V(Q,\omega) = A(Q)\delta(\omega) + B(Q)$. $A(Q)$ is proportional to the Debye-Waller factor, $\delta(\omega)$ is a Dirac function and $B(Q)$ is a background due to vibrations. $S_T(Q,\omega)$ can be represented by a single Lorentzian function $\mathcal{L}(\omega, \Gamma_T(Q))$ of half width at half maximum (HWHM) $\Gamma_T(Q)$. $S_R(Q,\omega)$ is expressed following the well-known Sears formalism:[49,50]

$$S_R(Q,\omega) = j_0^2(Qa)\delta(\omega)$$
$$+3j_1^2(Qa)\mathcal{L}\left(\omega, \frac{\hbar}{3\tau_R}\right)$$
$$+5j_2^2(Qa)\mathcal{L}\left(\omega, \frac{\hbar}{\tau_R}\right)$$

where $j_k$ is the k$^{th}$ Bessel function, $a$ is the radius of rotation and is taken to be the O-H distance in water molecule (0.98 Å), $\hbar$ is the reduced Planck constant and $\tau_R$ denotes the relaxation time of rotational diffusion. Considering the resolution of the instrument $R(\omega)$ and the above equations, the QENS signal of water $I_{water}(Q,\omega)$ can be rewritten as follow:

$$I_{water}(Q,\omega) = S_{water}(Q,\omega) \otimes R(\omega)$$
$$= A(Q)\left\{\begin{pmatrix} j_0^2(Qa)L(\omega, \Gamma_T(Q)) \\ +3j_1^2(Qa)L\left(\omega, \Gamma_T(Q) + \frac{\hbar}{3\tau_R}\right) \\ +5j_2^2(Qa)L\left(\omega, \Gamma_T(Q) + \frac{\hbar}{\tau_R}\right) \end{pmatrix} \otimes R(\omega)\right\}$$
$$+B(Q)$$

Within the random-jump-diffusion model,[51] the Q-dependence of $\Gamma_T$ is given as follow:

$$\Gamma_T(Q) = \frac{D_T Q^2}{1 + D_T \tau_0 Q^2}$$

where $D_T$ and $\tau_0$ are the translational diffusion constant and the residence time of the translational diffusion, respectively. In the low-Q limit, $\Gamma_T(Q)$ can be approximated by $D_T Q^2$ while in the high-Q limit, $\Gamma_T(Q)$ is approximately equal to $\tau_0^{-1}$.

We fit the QENS spectra of water using the above described model. $A(Q)$, $B(Q)$, $\Gamma_T(Q)$ and $\tau_R$ are fitted for both wavelengths, 5 Å and 8 Å, and for all the Qs. $\tau_R$ is shared through the entire dataset and $\Gamma_T(Q)$ are shared between the two wavelengths for the overlapping Q-range. The results of the fits are presented in Figure 7. The adopted model fits well both H$_2$O and D$_2$O. The Q-dependence of the extracted HWHM for H$_2$O follows the expected model behavior, and the extracted values for $\tau_R$, $D_T$ and $\tau_0$ are in a good agreement with the literature (see Table 2). Although the same behavior is observed for D$_2$O for Q-values up to 1.4 Å$^{-1}$, a different trend is observed for higher Qs, because D$_2$O presents a strong Bragg peak centered at higher Q, around ~ 1.8 Å$^{-1}$. Thus, the corresponding QENS spectra cannot be properly fitted and are therefore discarded.

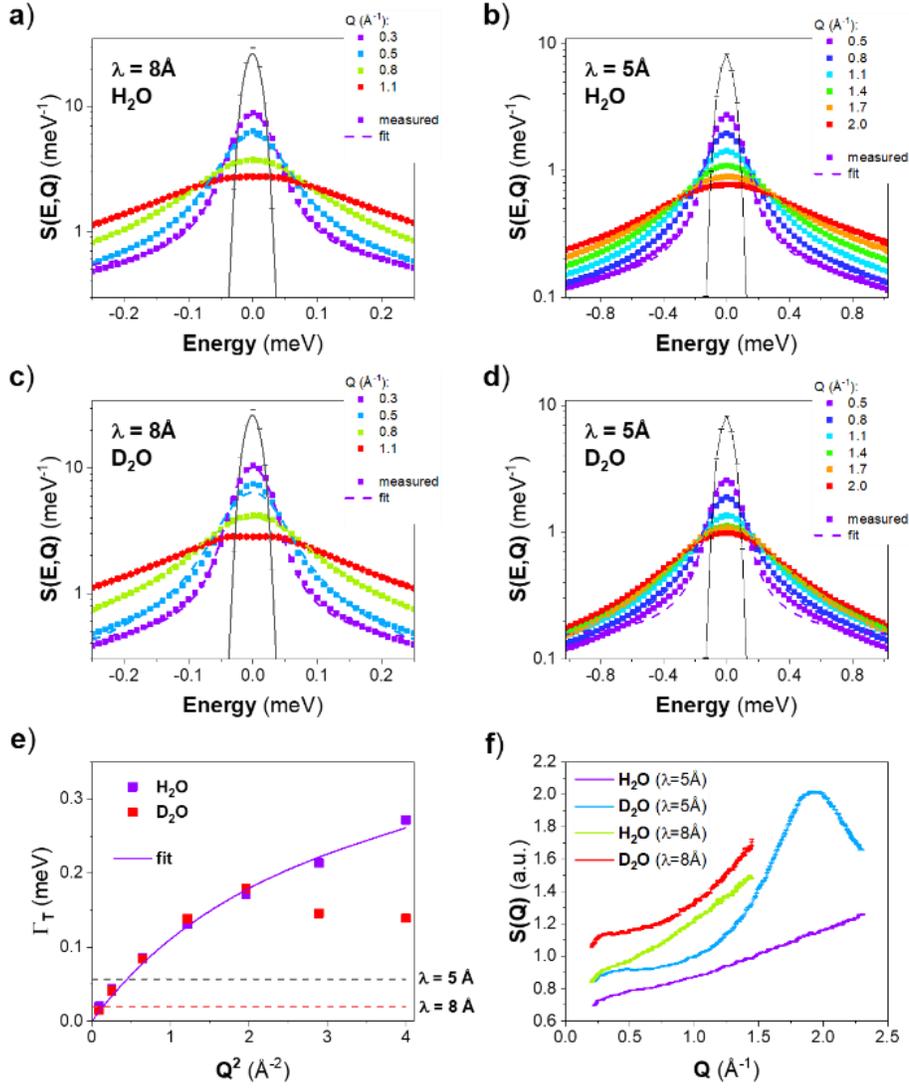

FIGURE 7. (a-d) Area-normalized QENS spectra (scatters), and their respective fit (dashed line), of $H_2O$ (a-b) and $D_2O$ (c-d), using two neutron incident wavelengths (two different instrumental resolutions) (a,c) $\lambda = 8$ Å and (b,d) $\lambda = 5$ Å. (e) HWHM, extracted from the fit of the QENS spectra of $H_2O$ and $D_2O$, as a function of $Q^2$, and the corresponding fit using the random-jump-diffusion model. The fit is done simultaneously for both the wavelengths and for all the Q values. The horizontal dashed lines represent the instrumental energy resolutions at $\lambda = 8$ Å and $\lambda = 5$ Å. (f) Neutron Diffractograms of $H_2O$ and $D_2O$ extracted from the measurements using the two wavelengths.

We fit the QENS signals of the two CTF materials using the same model for water to account for both constrained and free water, and by adding an extra contribution for the CTF as follow:

$$I(Q,\omega) = \left(S_{CTF}(Q,\omega) + S_{water}(Q,\omega)\right) \otimes R(\omega)$$
$$= C \times I_{CTF}(Q,\omega)$$
$$+ (1-C) \times A(Q) \times \left\{ \begin{pmatrix} j_0^2(Qa)L(\omega,\Gamma_T) \\ +3j_1^2(Qa)L\left(\omega,\Gamma_T(Q) + \frac{\hbar}{3\tau_R}\right) \\ +5j_2^2(Qa)L\left(\omega,\Gamma_T(Q) + \frac{\hbar}{\tau_R}\right) \end{pmatrix} \otimes R(\omega) \right\}$$
$$+ B(Q)$$

where $C$ is a shared parameter reflecting the concentration of water in the sample, weighted by the neutron incoherent cross-sections. Thus, it can, in principle, be calculated from Table 1. However, the presence of bound water can lead to an extra elastic contribution. The QENS signals of the CTFs exhibit a stronger elastic component than for bulk water and thus, the difference between $C$ extracted from the fit and calculated from Table 1 can be used to derive an estimate of the amount of bound water. No significant differences in the neutron diffractograms of the CTFs are observed upon addition of water (see Supporting Information Figure S1) $A(Q)$ is fixed here and taken to be equal to the values extracted from the fit to the free water (see Supporting Information Table S1).

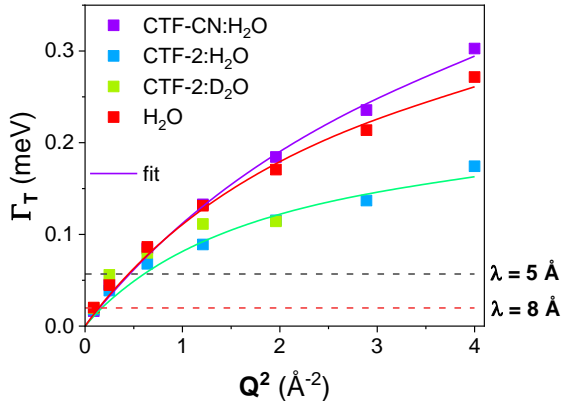

FIGURE 8. HWHM, extracted from the fit of the QENS spectra (see Supporting Information Figure S2), as a function of $Q^2$, and the corresponding fit using the random-jump-diffusion model for $H_2O$ and wetted CTFs: CTF-CN:$H_2O$, CTF-2:$H_2O$ and CTF-2:$D_2O$. The horizontal dashed lines represent the instrumental energy resolutions at $\lambda = 8$ Å and $\lambda = 5$ Å. The QENS spectra of CTF-CN:$D_2O$ are not fitted as we observed no significant differences between the QENS spectra of CTF-CN and CTF-CN:$D_2O$ (Figure 4 a,b).

Figure 8 presents the HWHM of the Lorentzian fits, of the QENS spectra, that accounts for the translational motion of water ($\Gamma_T$) for the different samples presently studied, except CTF-CN:$D_2O$. Surprisingly, no differences between $\Gamma_T$ of free water and $\Gamma_T$ of water in CTF-CN is observed within the error of the experiment/fit while water in CTF-2 appears to be constrained with a slower translational diffusion and a longer residence time $\tau_T$ (Table 2). For both CTFs, about 10 wt. % of water (Table 2) is either subjected to too slow motions, not captured by the instrumental time window, or is bound to the CTFs. This amount is seemingly slightly higher in the case of CTF-CN.

We could fit the CTF-2:$D_2O$ sample at lower Qs (the higher Qs being impacted by a strong Bragg peak of $D_2O$) and the extracted HWHM overlaps with the one extracted from the CTF-2:$H_2O$. It can be concluded from the very small change in the elastic peak between CTF-2 and CTF-2:$D_2O$ (Figure 6 c,d) combined with the HWHM overlap that almost no differences in dynamics is observed on the covered energy range for CTF-2. The differences observed in Figure 6 c,d are mainly due to the presence of constrained $D_2O$ and to the higher concentration of $D_2O$ used than for CTF-CN.

TABLE 2. Main fitting parameters for $H_2O$ and values from literature.[48] More parameters can be found in the Supporting Information (Table S1-2).

| | C | | Bound water | $\tau_R$ | $D_T$ | $\tau_T$ | $\chi^2$ |
| --- | --- | --- | --- | --- | --- | --- | --- |
| | expected | from fit | (wt. %) | (ps) | ($10^{-5}$cm$^2$ s$^{-1}$) | (ps) | |
| from ref.[48] | | | | 1.10 | 2.3 | 1.10 | |
| $H_2O$ | | | | 0.940 | 2.279 | 1.425 | 5.36 |
| CTF-CN:$H_2O$ (21.0 wt. % $H_2O$) | 0.533 | 0.778 | 12.3 | 1.316 | 2.120 | 1.056 | 10.24 |
| CTF-2:$H_2O$ (33.3 wt. % $H_2O$) | 0.408 | 0.531 | 9.4 | 1.296 | 1.958 | 2.752 | 15.59 |

DISCUSSION

The shorter strut of CTF-2 renders the structure stiffer. The softer structure of CTF-CN as measured by INS is correlated with a more pronounced dynamics on the 10s of picosecond timescale, as probed by QENS. The CTF-CN pore structure is thus more likely to collapse and this can be correlated with the smaller SA$_{BET}$.

We observed a guest/host type of dynamics for (wetted CTF-CN) CTF-CN:water. Further softening of the CTF-CN structure is observed by INS upon addition of water, while the libration degrees-of-freedom of water in CTF-CN are clearly hindered as shown by the changes in the rock, wag and twist modes of water within CTF-CN. No significant changes are observed in the case of CTF-2. The dynamics of both CTF-CN and CTF-2 on a longer time scale (10s of picoseconds) as measured by QENS is unchanged upon hydration. Moreover, more pronounced changes in the low-energy feature of water in CTF-CN is observed by INS. This indicates that a more pronounced change in the organization of water is occurring in the hydration monolayer of CTF-CN, leading to the hindrance of the libration of water in the hydration monolayer of CTF-CN. The hindrance of the librational degrees-of-freedom of water and the strong reorganization of water in CTF-CN is a clear indication of the transition of water from a free state to a bound state. From our QENS analysis, we found the amount of bound water in both CTF-CN and CTF-2 to be about 12 wt% and 9 wt%, respectively. No further constraint of water is observed for CTF-CN, while the translational diffusion of CTF-2 is clearly hindered in the case of CTF-2. It is worth noting that (i) our model does not discriminate between bound water and water with motions too slow to be captured by the instrumental time window, and (ii) the diffusion coefficient estimated from our model is an effective average between the diffusion coefficients of constrained water and free water.

In light of the above observations, it is reasonable to conclude that water and CTFs interact through two different mechanisms in CTF-CN and CTF-2. CTF-CN exhibits smaller SA$_{BET}$ and water uptake as the pore structure is softer. However, the -CN group interacts locally with water leading to a large amount of bound water and a strong rearrangement of the water hydration monolayer. This gives rise to a guest/host dynamics-type in CTF-CN. The water translational diffusion is not impacted in the case of CTF-CN, probably due to the presence of larger pores (mesopores), as found from the extracted pore size distribution. In CTF-2, the water diffusion is impacted as expected by the micro-porosity. The bound water observed by QENS is likely to be water diffusing too slowly to be captured by the instrumental resolution, as no strong evidence of bound water is observed by INS.

CONCLUSIONS

Covalent triazine-based frameworks (CTFs) are potential organic photocatalyst materials for water-splitting. The strategies to boost the rate of hydrogen production in organic photocatalysts have been so far to extend the overlap with the solar spectrum and reduce the thermodynamic driving force as well as speed up the kinetics of the multielectron redox reactions by using sacrificial agents and co-catalysts. However, rational design is still elusive as there is still a lack of an in-depth understanding of the processes occurring in a solvated catalyst.[52] The aqueous medium has been shown to ease the dissociation of the diffused excitons at the interface between the organic photocatalyst and water due to the higher relative permittivity, but this also means that the degree of solvation of the photocatalyst as well as the organization of the local environment, its permittivity and polarity when using sacrificial agent, can impact the thermodynamic force as well as the kinetics of the reactions. This is more striking for porous materials, with porosity that should be beneficial to increase the photocatalyst surface area, but does not always lead to higher activities.[36] Understanding the link between chemical design and local environment is therefore necessary.

Studying the microscopic interaction of water with the CTFs and the dynamics of the related mass transport within CTFs is of an utmost importance to help better characterizing CTFs properties with targeted tunable applications. In this context, we studied two CTFs; CTF-2 and CTF-CN, presenting structural differences in terms of the length and the polarity of their struts. CTF-CN has a longer strut (3 benzene rings against 2 for CTF-2) and has a polar -CN group. CTF-2 and CTF-CN exhibit similar electron affinities, ionization potentials, optical gap and they both swell. Surprisingly, the least performing CTF, CTF-2, presents the largest SA$_{BET}$ and water uptake. At first, this seems counterintuitive as larger SA$_{BET}$ and larger water uptake are believed to be beneficial to water-splitting, since increasing the interfacial surface area. However, previous modelling of oligomers with polar group suggested that the local environment of the polar group becomes more polar, improving the driving force for the oxidation half reaction. It appears that macroscopic measurements such as water uptake are not sufficient to explain the mechanism by which the -CN group leads to better photocatalytic water-splitting.

In this paper, we presented a detailed neutron spectroscopy study of CTF-2 and CTF-CN allowing us to probe and quantify the local water environment (bound water) and the related diffusion within the pores. Indeed, beyond the thermodynamic properties such as driving forces, the kinetics of the photocatalytic water-splitting process can be impacted by microporosity. Microporosity is likely to reduce the charge transport rates in the solid state and constrain the diffusion of both reactants and products in the pore. Here, we found that the -CN group promotes bound water and a strong rearrangement of the water hydration monolayer. However, the water translational diffusion is not impacted. In CTF-2, water diffusion is impacted as expected by the micro-porosity. Water dynamics should be balanced with the rate of reactions at the catalyst surface, which is still required to be studied. Nevertheless, it appears that further improvement such as engineering a stiffer pore structure could lead to a larger water uptake in the case of CTF-CN and thus, a larger photocatalytic activity, assuming that the diffusion of water in the pores balances the charge transport.

The design principle revealed in this study is that the underlying atomistic mode of interaction of water is important. Although previously suggested by theory, this is the first validation by experiment. In order to design better and more efficient CTF photocatalyst, the atomistic details of the polymer-water interactions need to be understood. Such systematic studies of chemical design, porosity and water dynamics can help defining chemical design rules or identifying the functional groups to be introduced to promote a "favorable" environment in terms of intermolecular interactions and dynamics. Therefore, the present work paves the way for a deeper understanding of the possible kinetic bottleneck limiting the efficiency of CTFs for water splitting applications.

EXPERIMENTAL SECTION

The neutron scattering measurements were performed on the direct geometry, cold neutron, disc chopper time-of-flight (TOF) spectrometer IN5 at the Institut Laue Langevin (Grenoble, France). An optimized sample thickness of 0.2 mm was considered, relevant to the minimization of effects like multiple scattering and absorption. Data were collected at 300K using two incident neutron wavelengths 5 Å ($E_i$ ≈ 3.27 meV) and 8Å ($E_i$ ≈ 1.28 meV), offering an optimal energy resolution at the elastic line of ~0.1 and 0.03 meV, respectively. We took the advantage offered by TOF-based neutron spectroscopy technique to extract both QENS and INS spectra, in addition to diffraction patterns, from the same acquired spectra. Indeed, the TOF data are acquired massively in time (energy) and space (momentum) transfer, thanks to a robust chopper system and a large detector/angular coverage, respectively, allowing to capture both elastic (detected neutrons without energy exchange with the sample), and quasi-elastic and inelastic processes (detected neutrons having exchanged an amount of energy with the sample). Standard corrections including detector efficiency calibration and background subtraction were performed. A vanadium sample was used to calibrate the detectors and to measure the instrumental resolution under the same operating conditions. At the wavelengths used, the IN5 angular detector coverage corresponds to a Q-range of ~ 0.2–2.3 Å$^{-1}$ ($\lambda_i$= 5Å) and ~ 0.1–1.3 Å$^{-1}$ ($\lambda_i$ = 8Å). The data reduction and analysis were done using ILL software tools[53] to extract the neutron diffraction patterns, the QENS spectra and the generalized density of states (GDOS). The neutron diffractograms were extracted by averaging the energy around the elastic peak of the scattering function S(Q,E). For the QENS spectra, different data sets were extracted either by performing a full Q-average in the (Q, E) space to get the scattering function S(E, T) or by considering Q-slices to study the S(Q, E, T). The presented one-phonon GDOS,[45,54] spectra were extracted from the inelastic part of the S(Q,E), using the neutron incident wavelength of 5 Å, and operating in the up-scattering, neutron energy-gain mode.

## ASSOCIATED CONTENT

**Supporting Information**. Neutron diffractograms, QENS spectra and associated fits with supplementary fitting parameters, Pore size distribution. This material is available free of charge via the Internet at http://pubs.acs.org.

## AUTHOR INFORMATION

### Corresponding Author

* zbiri@ill.fr
* a.guilbert09@imperial.ac.uk

### Author Contributions

A.A.Y.G. and M.Z. conceived and developed the project, planned and carried out the neutron measurements, treated and analyzed the neutron data. C.M.A., R.S.S. and A.I.C. synthesized the materials and performed the associated characterization. A.A.Y.G. and M.Z. wrote the manuscript with contribution from coauthors.

## ACKNOWLEDGMENT

The Institut Laue-Langevin (ILL) facility (Grenoble, France) is acknowledged for providing beam time on the IN5 spectrometer. A.A.Y.G. acknowledges the Engineering and Physical Sciences Research Council (EPSRC) for the award of an EPSRC Postdoctoral Fellowship (EP/P00928X/1). C.M.A, A.I.C and R.S.S thank EPSRC for the financial support under Grant EP/N004884/1. Rob Clowes is thanked for help with the sorption measurements.

## REFERENCES

(1) Zhao, Y.; Ding, C.; Zhu, J.; Qin, W.; Tao, X.; Fan, F.; Li, R.; Li, C. A Hydrogen Farm Strategy for Scalable Solar Hydrogen Production with Particulate Photocatalysts. *Angew. Chemie* **2020**, *132* (24), 9740–9745. https://doi.org/10.1002/ange.202001438.
(2) Takata, T.; Domen, K. Particulate Photocatalysts for Water Splitting: Recent Advances and Future Prospects. *ACS Energy Lett.* **2019**, *4* (2), 542–549. https://doi.org/10.1021/acsenergylett.8b02209.
(3) Wang, Z.; Li, C.; Domen, K. Recent Developments in Heterogeneous Photocatalysts for Solar-Driven Overall Water Splitting. *Chem. Soc. Rev.* **2019**, *48* (7), 2109–2125. https://doi.org/10.1039/c8cs00542g.
(4) Jayakumar, J.; Chou, H. H. Recent Advances in Visible-Light-Driven Hydrogen Evolution from Water Using Polymer Photocatalysts. *ChemCatChem* **2020**, *12* (3), 689–704. https://doi.org/10.1002/cctc.201901725.
(5) Zhang, G.; Lan, Z.-A.; Wang, X. Conjugated Polymers: Catalysts for Photocatalytic Hydrogen Evolution. *Angew. Chemie Int. Ed.* **2016**, *55* (51), 15712–15727. https://doi.org/10.1002/anie.201607375.
(6) Wang, Q.; Hisatomi, T.; Jia, Q.; Tokudome, H.; Zhong, M.; Wang, C.; Pan, Z.; Takata, T.; Nakabayashi, M.; Shibata, N.; Li, Y.; Sharp, I. D.; Kudo, A.; Yamada, T.; Domen, K. Scalable Water Splitting on Particulate Photocatalyst Sheets with a Solar-to-Hydrogen Energy Conversion Efficiency Exceeding 1%. *Nat. Mater.* **2016**, *15* (6), 611–615. https://doi.org/10.1038/nmat4589.
(7) Wang, X.; Zhang, G.; Lan, Z.-A. Organic Conjugated Semiconductors for Photocatalytic Hydrogen Evolution with Visible Light. *Angew. Chemie, Int. Ed.* **2016**, *55* (51), 15712–15727. https://doi.org/10.1002/anie.201607375.
(8) Wang, X.; Maeda, K.; Thomas, A.; Takanabe, K.; Xin, G.; Carlsson, J. M.; Domen, K.; Antonietti, M. A Metal-Free Polymeric Photocatalyst for Hydrogen Production from Water under Visible Light. *Nat. Mater.* **2009**, *8* (1), 76–80. https://doi.org/10.1038/nmat2317.
(9) Kong, D.; Zheng, Y.; Kobielusz, M.; Wang, Y.; Bai, Z.; Macyk, W.; Wang, X.; Tang, J. Recent Advances in Visible Light-Driven Water Oxidation and Reduction in Suspension Systems. *Mater. Today* **2018**, *21* (8), 897–924. https://doi.org/10.1016/j.mattod.2018.04.009.
(10) Yanagida, S.; Kabumoto, A.; Mizumoto, K.; Pac, C.; Yoshino, K. Poly(P-Phenylene)-Catalysed Photoreduction of Water to Hydrogen. *J. Chem. Soc. - Ser. Chem. Commun.* **1985**, No. 8, 474–475. https://doi.org/10.1039/C39850000474.
(11) Matsuoka, S.; Kohzuki, T.; Nakamura, A.; Pac, C.; Yanagida, S. Efficient Visible-Light-Driven Photocatalysis. Poly(Pyridine-2,5-Diyl)-Catalysed Hydrogen Photoevolution and Photoreduction of Carbonyl Compounds. *J. Chem. Soc. Chem. Commun.* **1991**, *39* (8), 580. https://doi.org/10.1039/c39910000580.
(12) Sprick, R. S.; Bonillo, B.; Clowes, R.; Guiglion, P.; Brownbill, N. J.; Slater, B. J.; Blanc, F.; Zwijnenburg, M. A.; Adams, D. J.; Cooper, A. I. Visible-Light-Driven Hydrogen Evolution Using Planarized Conjugated Polymer Photocatalysts. *Angew. Chemie Int. Ed.* **2016**, *55* (5), 1792–1796. https://doi.org/10.1002/anie.201510542.
(13) Sprick, R. S.; Aitchison, C. M.; Berardo, E.; Turcani, L.; Wilbraham, L.; Alston, B. M.; Jelfs, K. E.; Zwijnenburg, M. A.; Cooper, A. I. Maximising the Hydrogen Evolution Activity in Organic Photocatalysts by Co-Polymerisation. *J. Mater. Chem. A* **2018**, *6* (25), 11994–12003. https://doi.org/10.1039/C8TA04186E.
(14) Kosco, J.; Bidwell, M.; Cha, H.; Martin, T.; Howells, C. T.; Sachs, M.; Anjum, D. H.; Gonzalez Lopez, S.; Zou, L.; Wadsworth, A.; Zhang, W.; Zhang, L.; Tellam, J.; Sougrat, R.; Laquai, F.; DeLongchamp, D. M.; Durrant, J. R.; McCulloch, I. Enhanced Photocatalytic Hydrogen Evolution from Organic Semiconductor Heterojunction Nanoparticles. *Nat. Mater.* **2020**, *19*, 559–565. https://doi.org/10.1038/s41563-019-0591-1.
(15) Sprick, R. S.; Cheetham, K. J.; Bai, Y.; Alves Fernandes, J.; Barnes, M.; Bradley, J. W.; Cooper, A. I. Polymer Photocatalysts with Plasma-Enhanced Activity. *J. Mater. Chem. A* **2020**, *8*, 7125–7129. https://doi.org/10.1039/D0TA01200A.
(16) Tseng, P.-J. J.; Chang, C.-L. L.; Chan, Y.-H. H.; Ting, L.-Y. Y.; Chen, P.-Y. Y.; Liao, C.-H. H.; Tsai, M.-L. L.; Chou, H.-H. H. Design and Synthesis of Cycloplatinated Polymer Dots as Photocatalysts for Visible Light-Driven Hydrogen Evolution. *ACS Catal.* **2018**, *8* (9), 7766–7772. https://doi.org/10.1021/acscatal.8b01678.
(17) Lin, K.; Wang, Z.; Hu, Z.; Luo, P.; Yang, X.; Zhang, X.; Rafiq, M.;

Huang, F.; Cao, Y. Amino-Functionalised Conjugated Porous Polymers for Improved Photocatalytic Hydrogen Evolution. *J. Mater. Chem. A* **2019**, *7* (32), 19087–19093. https://doi.org/10.1039/c9ta06219j.

(18) Yu, J.; Sun, X.; Xu, X.; Zhang, C.; He, X. Donor-Acceptor Type Triazine-Based Conjugated Porous Polymer for Visible-Light-Driven Photocatalytic Hydrogen Evolution. *Appl. Catal. B Environ.* **2019**, *257*, 117935. https://doi.org/https://doi.org/10.1016/j.apcatb.2019.117935.

(19) Sprick, R. S.; Jiang, J.-X.; Bonillo, B.; Ren, S.; Ratvijitvech, T.; Guiglion, P.; Zwijnenburg, M. A.; Adams, D. J.; Cooper, A. I. Tunable Organic Photocatalysts for Visible-Light-Driven Hydrogen Evolution. *J. Am. Chem. Soc.* **2015**, *137* (9), 3265–3270. https://doi.org/10.1021/ja511552k.

(20) Li, L.; Lo, W. Y.; Cai, Z.; Zhang, N.; Yu, L. Donor-Acceptor Conjugated Polymers for Photocatalytic Hydrogen Production: The Importance of Acceptor Comonomer. *Macromolecules* **2016**, *49* (18), 6903–6909. https://doi.org/10.1021/acs.macromol.6b01764.

(21) Li, L.; Cai, Z.; Wu, Q.; Lo, W. Y.; Zhang, N.; Chen, L. X.; Yu, L. Rational Design of Porous Conjugated Polymers and Roles of Residual Palladium for Photocatalytic Hydrogen Production. *J. Am. Chem. Soc.* **2016**, *138* (24), 7681–7686. https://doi.org/10.1021/jacs.6b03472.

(22) Bi, S.; Lan, Z.; Paasch, S.; Zhang, W.; He, Y.; Zhang, C.; Liu, F.; Wu, D.; Zhuang, X.; Brunner, E.; Wang, X.; Zhang, F. Substantial Cyano-Substituted Fully Sp2-Carbon-Linked Framework: Metal-Free Approach and Visible-Light-Driven Hydrogen Evolution. *Adv. Funct. Mater.* **2017**, *27* (39). https://doi.org/10.1002/adfm.201703146.

(23) Kochergin, Y. S.; Schwarz, D.; Acharjya, A.; Ichangi, A.; Kulkarni, R.; Eliášová, P.; Vacek, J.; Schmidt, J.; Thomas, A.; Bojdys, M. J. Exploring the "Goldilocks Zone" of Semiconducting Polymer Photocatalysts by Donor-Acceptor Interactions. *Angew. Chemie Int. Ed.* **2018**, *57* (43), 14188–14192. https://doi.org/10.1002/anie.201809702.

(24) Yang, C.; Ma, B. C.; Zhang, L.; Lin, S.; Ghasimi, S.; Landfester, K.; Zhang, K. A. I.; Wang, X. Molecular Engineering of Conjugated Polybenzothiadiazoles for Enhanced Hydrogen Production by Photosynthesis. *Angew. Chemie - Int. Ed.* **2016**, *55* (32), 9202–9206. https://doi.org/10.1002/anie.201603532.

(25) Meier, C. B.; Sprick, R. S.; Monti, A.; Guiglion, P.; Lee, J.-S. M.; Zwijnenburg, M. A.; Cooper, A. I. Structure-Property Relationships for Covalent Triazine-Based Frameworks: The Effect of Spacer Length on Photocatalytic Hydrogen Evolution from Water. *Polymer (Guildf).* **2017**, *126*, 283–290. https://doi.org/10.1016/j.polymer.2017.04.017.

(26) Meier, C. B.; Clowes, R.; Berardo, E.; Jelfs, K. E.; Zwijnenburg, M. A.; Sprick, R. S.; Cooper, A. I. Structurally Diverse Covalent Triazine-Based Framework Materials for Photocatalytic Hydrogen Evolution from Water. *Chem. Mater.* **2019**, *31* (21), 8830–8838. https://doi.org/10.1021/acs.chemmater.9b02825.

(27) Xie, J.; Shevlin, S. A.; Ruan, Q.; Moniz, S. J. A.; Liu, Y.; Liu, X.; Li, Y.; Lau, C. C.; Guo, Z. X.; Tang, J. Efficient Visible Light-Driven Water Oxidation and Proton Reduction by an Ordered Covalent Triazine-Based Framework. *Energy Environ. Sci.* **2018**, *11*, 1617–1624. https://doi.org/10.1039/C7EE02981K.

(28) Lan, Z.-A. A.; Fang, Y.; Zhang, Y.; Wang, X. Photocatalytic Oxygen Evolution from Functional Triazine-Based Polymers with Tunable Band Structures. *Angew. Chemie - Int. Ed.* **2018**, *57* (2), 470–474. https://doi.org/10.1002/anie.201711155.

(29) Wang, X.; Chen, L.; Chong, S. Y.; Little, M. A.; Wu, Y.; Zhu, W.-H.; Clowes, R.; Yan, Y.; Zwijnenburg, M. A.; Sprick, R. S.; Cooper, A. I. Sulfone-Containing Covalent Organic Frameworks for Photocatalytic Hydrogen Evolution from Water. *Nat. Chem.* **2018**, *10*, 1180–1189. https://doi.org/10.1038/s41557-018-0141-5.

(30) Banerjee, T.; Haase, F.; Savasci, G.; Gottschling, K.; Ochsenfeld, C.; Lotsch, B. V. Single-Site Photocatalytic H2 Evolution from Covalent Organic Frameworks with Molecular Cobaloxime Co-Catalysts. *J. Am. Chem. Soc.* **2017**, *139* (45), 16228–16234. https://doi.org/10.1021/jacs.7b07489.

(31) Vyas, V. S.; Haase, F.; Stegbauer, L.; Savasci, G.; Podjaski, F.; Ochsenfeld, C.; Lotsch, B. V. A Tunable Azine Covalent Organic Framework Platform for Visible Light-Induced Hydrogen Generation. *Nat. Commun.* **2015**, *6*, 8508. https://doi.org/10.1038/ncomms9508.

(32) Pachfule, P.; Acharjya, A.; Roeser, J.; Langenhahn, T.; Schwarze, M.; Schomäcker, R.; Thomas, A.; Schmidt, J. Diacetylene Functionalized Covalent Organic Framework (COF) for Photocatalytic Hydrogen Generation. *J. Am. Chem. Soc.* **2018**, *140* (4), 1423–1427. https://doi.org/10.1021/jacs.7b11255.

(33) Jin, E.; Lan, Z.; Jiang, Q.; Geng, K.; Li, G.; Wang, X.; Jiang, D. 2D Sp2 Carbon-Conjugated Covalent Organic Frameworks for Photocatalytic Hydrogen Production from Water. *Chem* **2019**, *5* (6), 1632–1647. https://doi.org/10.1016/j.chempr.2019.04.015.

(34) Sachs, M.; Sprick, R. S.; Pearce, D.; Hillman, S. A. J.; Monti, A.; Guilbert, A. A. Y.; Brownbill, N. J.; Dimitrov, S.; Shi, X.; Blanc, F.; Zwijnenburg, M. A.; Nelson, J.; Durrant, J. R.; Cooper, A. I. Understanding Structure-Activity Relationships in Linear Polymer Photocatalysts for Hydrogen Evolution. *Nat. Commun.* **2018**, *9* (1), 4968. https://doi.org/10.1038/s41467-018-07420-6.

(35) Woods, D. J.; Hillman, S. A. J.; Pearce, D.; Wilbraham, L.; Flagg, L. Q.; Duffy, W.; Mcculloch, I.; Durrant, J. R.; Guilbert, A. A. Y.; Zwijnenburg, M. A.; Sprick, R. S.; Nelson, J.; Cooper, A. I. Side-Chain Tuning in Conjugated Polymer Photocatalysts for Improved Hydrogen Production from Water. *Energy Environ. Sci.* **2020**, *13* (6), 1843–1855. https://doi.org/10.1039/d0ee01213k.

(36) Sprick, R. S.; Bai, Y.; Guilbert, A. A. Y.; Zbiri, M.; Aitchison, C. M.; Wilbraham, L.; Yan, Y.; Woods, D. J.; Zwijnenburg, M. A.; Cooper, A. I. Photocatalytic Hydrogen Evolution from Water Using Fluorene and Dibenzothiophene Sulfone-Conjugated Microporous and Linear Polymers. *Chem. Mater.* **2019**, *31* (2), 305–313. https://doi.org/10.1021/acs.chemmater.8b02833.

(37) Xu, Y.; Mao, N.; Feng, S.; Zhang, C.; Wang, F.; Chen, Y.; Zeng, J.; Jiang, J.-X. Perylene-Containing Conjugated Microporous Polymers for Photocatalytic Hydrogen Evolution. *Macromol. Chem. Phys.* **2017**, *218* (14), 1700049. https://doi.org/10.1002/macp.201700049.

(38) Guilbert, A. A. Y.; Zbiri, M.; Finn, P. A.; Jenart, M.; Fouquet, P.; Cristiglio, V.; Frick, B.; Nelson, J.; Nielsen, C. B. Mapping Microstructural Dynamics up to the Nanosecond of the Conjugated Polymer P3HT in the Solid State. *Chem. Mater.* **2019**, *31* (23), 9635–9651. https://doi.org/10.1021/acs.chemmater.9b02904.

(39) Wagemaker, M.; Kearley, G. J.; Van Well, A. A.; Mutka, H.; Mulder, F. M. Multiple Li Positions inside Oxygen Octahedra in Lithiated TiO2 Anatase. *J. Am. Chem. Soc.* **2003**, *125* (3), 840–848. https://doi.org/10.1021/ja028165q.

(40) Rosenbach, N.; Jobic, H.; Ghoufi, A.; Salles, F.; Maurin, G.; Bourrelly, S.; Llewellyn, P. L.; Devic, T.; Serre, C.; Férey, G. Quasi-Elastic Neutron Scattering and Molecular Dynamics Study of Methane Diffusion in Metal Organic Frameworks MIL-47(V) and MIL-53(Cr). *Angew. Chemie Int. Ed.* **2008**, *47* (35), 6611–6615. https://doi.org/10.1002/anie.200801748.

(41) Pham, T.; Forrest, K. A.; Mostrom, M.; Hunt, J. R.; Furukawa, H.; Eckert, J.; Space, B. The Rotational Dynamics of H 2 Adsorbed in Covalent Organic Frameworks. *Phys. Chem. Chem. Phys.* **2017**, *19* (20), 13075–13082. https://doi.org/10.1039/C7CP00924K.

(42) Ruffle, S. V.; Michalarias, I.; Li, J. C.; Ford, R. C. Inelastic Incoherent Neutron Scattering Studies of Water Interacting with Biological Macromolecules. *J. Am. Chem. Soc.* **2002**, *124* (4), 565–569. https://doi.org/10.1021/ja016277w.

(43) Liu, M.; Chen, L.; Lewis, S.; Chong, S. Y.; Little, M. A.; Hasell, T.;

(43) Aldous, I. M.; Brown, C. M.; Smith, M. W.; Morrison, C. A.; Hardwick, L. J.; Cooper, A. I. Three-Dimensional Protonic Conductivity in Porous Organic Cage Solids. *Nat. Commun.* **2016**, *7* (1), 12750. https://doi.org/10.1038/ncomms12750.

(44) P. Barrett, E.; G. Joyner, L.; P. Halenda, P. The Determination of Pore Volume and Area Distributions in Porous Substances. I. Computations from Nitrogen Isotherms. *J. Am. Chem. Soc.* **2002**, *73* (1), 373–380. https://doi.org/10.1021/ja01145a126.

(45) *A Generalized Density of States (GDOS) Is the Phonon Spectrum Measured from Inelastic Neutron Scattering (INS). In Contrast to the Vibrational Density of States, the GDOS Involves a Weighting of the Scatterers (Atoms) with Their Scattering Powers σ/M (σ: cross section, M: mass). These are presently for various atoms in the units of [barns·amu−1]: H, 81.37; D, 3.8; C, 0.46; and N, 0.822.*

(46) Cygan, R. T.; Daemen, L. L.; Ilgen, A. G.; Krumhansl, J. L.; Nenoff, T. M. Inelastic Neutron Scattering and Molecular Simulation of the Dynamics of Interlayer Water in Smectite Clay Minerals. *J. Phys. Chem. C* **2015**, *119* (50), 28005–28019. https://doi.org/10.1021/acs.jpcc.5b08838.

(47) Gong, K.; Cheng, Y.; Daemen, L. L.; White, C. E. In Situ Quasi-Elastic Neutron Scattering Study on the Water Dynamics and Reaction Mechanisms in Alkali-Activated Slags. *Phys. Chem. Chem. Phys.* **2019**, *21* (20), 10277–10292. https://doi.org/10.1039/c9cp00889f.

(48) Teixeira, J.; Bellissent-Funel, M.-C.; Chen, S. H.; Dianoux, A. J. Experimental Determination of the Nature of Diffusive Motions of Water Molecules at Low Temperatures. *Phys. Rev. A* **1985**, *31* (3), 1913–1917. https://doi.org/10.1103/PhysRevA.31.1913.

(49) Sears, V. F. THEORY OF COLD NEUTRON SCATTERING BY HOMONUCLEAR DIATOMIC LIQUIDS: I. FREE ROTATION. *Can. J. Phys.* **1966**, *44* (6), 1279–1297. https://doi.org/10.1139/p66-108.

(50) Sears, V. F. THEORY OF COLD NEUTRON SCATTERING BY HOMONUCLEAR DIATOMIC LIQUIDS: II. HINDERED ROTATION. *Can. J. Phys.* **1966**, *44* (6), 1299–1311. https://doi.org/10.1139/p66-109.

(51) Bee, M. Quasielastic Neutron Scattering for Continuous or Random Jump Diffusion of Molecules in Bounded Media. In *Quasielastic neutron scattering : principles and applications in solid state chemistry, biology, and materials science*; Adam Hilger: Bristol, Philadelphia, 1988; pp 357–398.

(52) Rahman, M.; Tian, H.; Edvinsson, T. Revisiting the Limiting Factors for Overall Water-Splitting on Organic Photocatalysts. *Angew. Chemie - Int. Ed.* **2020**, *59* (38), 16278–16293. https://doi.org/10.1002/anie.202002561.

(53) Richard, D.; Ferrand, M.; Kearley, G. J. Analysis and Visualisation of Neutron-Scattering Data. *J. Neutron Res.* **1996**, *4* (1–4), 33–39. https://doi.org/10.1080/10238169608200065.

(54) Sjölander, A. Multi-Phonon Processes in Slow Neutron Scattering by Crystals. *Ark. För Fys.* **1958**, *14* (21), 315–371.